\documentclass[sigconf]{acmart}

\usepackage{tikz}
\usetikzlibrary{shapes,positioning,arrows}
\usepackage{bbding}
\usepackage{xspace}

\newcommand{\cdcl}{CDCL(Crypto)\xspace}

\AtBeginDocument{%
  \providecommand\BibTeX{{%
    \normalfont B\kern-0.5em{\scshape i\kern-0.25em b}\kern-0.8em\TeX}}}

\acmConference[CASCON'19]{CASCON'19}{November 4--6, 2019}{Toronto, Ontario}

\begin{document}

\title{CDCL(Crypto) SAT Solvers for Cryptanalysis}

\author{Saeed Nejati}
\email{snejati@uwaterloo.ca}
\affiliation{%
  \institution{University of Waterloo}
  \streetaddress{200 University Ave W}
  \city{Waterloo}
  \state{Ontario}
  \postcode{N2L 3G1}
}

\author{Vijay Ganesh}
\email{vganesh@uwaterloo.ca}
\affiliation{%
  \institution{University of Waterloo}
  \streetaddress{200 University Ave W}
  \city{Waterloo}
  \state{Ontario}
  \postcode{N2L 3G1}
}

\renewcommand{\shortauthors}{Nejati and Ganesh}

\begin{abstract}
    Over the last two decades, we have seen a dramatic improvement in the efficiency of conflict-driven clause-learning Boolean satisfiability (CDCL SAT) solvers on industrial problems from a variety of domains. The availability of such powerful general-purpose search tools as SAT solvers has led many researchers to propose SAT-based methods for cryptanalysis, including techniques for finding collisions in hash functions and breaking symmetric encryption schemes. Most of the previously proposed SAT-based cryptanalysis approaches are {\it blackbox} techniques, in the sense that the cryptanalysis problem is encoded as a SAT instance and then a CDCL SAT solver is invoked to solve the said instance. A weakness of this approach is that the encoding thus generated may be too large for any modern solver to solve efficiently.  Perhaps a more important weakness of this approach is that the solver is in no way specialized or tuned to solve the given instance.
    To address these issues, we propose an approach called \cdcl (inspired by the CDCL(\textit{T}) paradigm in Satisfiability Modulo Theory solvers) to tailor the internal subroutines of the CDCL SAT solver with domain-specific knowledge about cryptographic primitives. Specifically, we extend the propagation and conflict analysis subroutines of CDCL solvers with specialized codes that have knowledge about the cryptographic primitive being analyzed by the solver. We demonstrate the power of this approach in the differential path and algebraic fault analysis of hash functions. Our initial results are very encouraging and reinforce the notion that this approach is a significant improvement over blackbox SAT-based cryptanalysis.
\end{abstract}




\maketitle

\section{Introduction}
Boolean satisfiability (SAT) solvers are well-known as powerful general purpose search tools, that have been used in solving problems from many different domains, such as verification, AI and cryptography \cite{cadar2008exe}, \cite{rintanen2009planning}, \cite{mironov2006applications}. They get their power from reasoning components like clause learning \cite{marques1999grasp} and many different search heuristics, like VSIDS or machine-learning based LRB branching \cite{moskewicz2001chaff}, \cite{liang2016learning} clause deletion \cite{audemard2009predicting} and restarts \cite{audemard2012refining}.

\textbf{SAT-based Cryptanalysis}.
The availability of such powerful search tool has led many researchers to propose the use of SAT and SMT solvers for cryptanalysis of hash functions and symmetric encryption schemes, for example in preimage attacks \cite{morawiecki2013sat}, \cite{nossum2012sat}, collision attacks \cite{mironov2006applications}, \cite{prokop2016differential} and linear and differential cryptanalysis of block ciphers\cite{ashur2017automated}, \cite{kolbl2015observations}.

Although in some of the approaches, the heuristics of the solver are altered to improve their efficiency, e.g. branching heuristics \cite{prokop2016differential}, \cite{semenov2011parallel} and restart policy \cite{nejati2017adaptive}, most of these approaches used a direct encoding of the said problems into a satisfiability problem and used SAT solvers as a blackbox, and the changes are limited to the search heuristics and do not alter the logic reasoning components of the solver. The one notable exception is the CryptoMiniSat solver \cite{soos2009extending}, that adds reasoning over XOR clauses to the solver to improve the solving of cryptographic instances that heavily use XOR operations.

The current work on SAT-based cryptanalysis is similar to the \textit{eager} approach in solving Satisfiability Modulo Theories (SMT) formulas, where the formula is directly translated into a SAT instance and then a SAT solver is invoked on it. The benefit of this approach is that we can use any SAT solver as-is and leverage the performance of the solver and its improvement capacity over time. The downside of this approach is the loss of the high level semantics of the underlying theories. This means that the SAT solver needs to perform extra computations to prove facts that are readily available in the higher level logic (e.g. $x+y = y+x$ in the integer arithmetic). The other main approach of solving SMT instances, called \textit{lazy} approach, integrates the CDCL style search with theory-specific solvers ($T$-solvers). This architecture is referred to as CDCL($T$). Generally speaking, a $T$-solver is useful only if it participates in propagation and conflict analysis reasoning engines of the SAT solver they extend.

\textbf{Our Contributions}. The main research question that we pose in this paper is: \\
\textit{Q: Are there methods that can surpass blackbox SAT-based cryptanalysis in terms of scalability and ability to break complex real-world cryptographic primitives?} \\
\textbf{1. } Inspired by the CDCL($T$) paradigm, we propose a framework for SAT-based cryptanalysis that we call \cdcl. It extends the propagation and conflict analysis of the core SAT solver using the higher level knowledge about the cryptographic problem that is being analyzed. To be more flexible, and to have simpler implementation and be able to customize the extended functionalities to different cryptographic problems, we use the \textit{Programmatic SAT} \cite{ganesh2012lynx} architecture, where the solver provides callbacks for extending propagation and conflict analysis to be implemented by the user.

\textbf{2. } We first review an application of this framework that has been successfully applied to algebraic fault analysis of SHA-1 and SHA-256 cryptographic hash functions \cite{nejati2018algebraic}, enabling the attacker to recover the secret bits with only 11 faults in SHA-1 and 48 faults in SHA-256, which is a significant improvement over previous algebraic fault attacks. 

\textbf{3. } Then we demonstrate that this framework can be applied to other cryptographic problems, more specifically differential cryptanalysis of round reduced SHA-256. We present preliminary results on increasing the number of rounds in the collision finding of SHA-256 compared to the previous SAT-based differential cryptanalysis of SHA-256.
 

\section{Preliminaries}
We refer the reader to the Handbook of Satisfiability~\cite{biere2009handbook} for a detailed discussion on SAT and SMT solvers.

\subsection{Algebraic Fault Attack}
Fault attack is an invasive attack on the implementation of a cryptographic primitive that has an embedded message or secret key. The attacker expresses the set of cryptographic relations in an algebraic setting with the correct output value. Then the attacker runs the function again but this time induces faults during the process of the function. This fault changes the value of a targeted register and causes the function to output a faulty value. The attacker then expresses the same set of relations but with the faulty output. These additional equations will constrain the size of possible values for the secret bits. The fault injecting process can be performed multiple times to obtain a more constrained equation set but at the cost of formula size and solver effort. The fault is usually injected in the input of a round close to the output, to study the propagation of information through a small number of rounds. The equation set for a function $f = f_1 o f_2$ will be $y = f_1(f_2(x)) \land y'_1 = f_1(\delta_1 \oplus f_2(x)) \land y'_2 = f_1(\delta_2 \oplus f_2(x))$, where $y'_i$s are faulty outputs and $\delta_i$s are random fault values that are unknown to the attacker. This equation set is then handed over to a SAT/SMT solver to find the secret bits. The equation set corresponding to the rounds from the beginning of the function to the point of fault injection  (equations for $f_2(x)$), are the same among correct and faulty outputs and provide the same value to the second part of the equations. Therefore it is a common practice to remove these equations altogether and focus on the fault injected part of the function. But because this is an abstraction of the full function, the found solution needs to be verified against the full version.

\subsection{Differential Cryptanalysis}
Broadly speaking, differential cryptanalysis \cite{biham1991differential} is the analysis of how a difference in the input values of a cryptographic function can affect the resultant difference at the output. Block ciphers and cryptographic hash functions are typically comprised of chaining of smaller functions. In these cases, differential cryptanalysis looks at the trace of differences of values through the chain of transformations to find non-random behaviors of the function and exploiting it to find input messages or secret keys.

Usually XOR difference is considered ($\Delta x = x \oplus x'$). We are interested in relations between $\Delta x$ and $\Delta y = f(x \oplus \Delta x) \oplus f(x)$, for a cryptographic function $f$. Describing and analyzing the differential ($\Delta x \rightarrow \Delta y$) over $f$ itself is usually impractical. Therefore the differentials over the smaller steps of $f$ are analyzed and chained together to derive differentials over input/outputs of $f$. The combination of this chain of differentials over smaller operations is called a differential path or trace. To express the set of possible combinations of a pair of bits $x$ and $x'$, the generalized conditions \cite{de2006finding} are commonly used. It allows us to describe and encode the propagation of information through a differential path. This notation is listed in table \ref{tab:diffnotation}.

\begin{table}[ht]
    \caption{Notation for all generalized conditions. Each character represents the set of possible
    values for a pair of bits.}
    \bgroup
    \setlength{\tabcolsep}{4pt}
    \begin{tabular}{|c|cccccccccccccccc|} \hline
        $(x_i, x_i')$ & 
        \texttt{?} & \texttt{-} & \texttt{x} & \texttt{0} & \texttt{u} & \texttt{n} & \texttt{1} & \texttt{\#} &
        \texttt{3} & \texttt{5} & \texttt{7} & \texttt{A} & \texttt{B} & \texttt{C} & \texttt{D} & \texttt{E} \\
        \hline
        (0, 0) & + & + &   & + &   &   &   &   & + & + & + &   & + &   & + &   \\
        (1, 0) & + &   & + &   & + &   &   &   & + &   & + & + & + &   &   & + \\
        (0, 1) & + &   & + &   &   & + &   &   &   & + & + &   &   & + & + & + \\
        (1, 1) & + & + &   &   &   &   & + &   &   &   &   & + & + & + & + & + \\
        \hline
    \end{tabular}
    \egroup
    \label{tab:diffnotation}
\end{table}

\begin{figure*}[t]
    \centering
    \resizebox{16.5cm}{!}{
    \begin{tikzpicture}
    \tikzset{>=latex'}
    \tikzset{auto}
    \tikzstyle{block} = [draw, rectangle, rounded corners, minimum height=3em, minimum width=6em]
    \tikzstyle{decision} = [diamond, draw, text badly centered, inner sep=1pt]
    
    \node (input) {Input Formula};
    \node [block, below of=input, node distance=1.2cm] (unit) {Unit Propagation};
    \node [decision, below of=unit, node distance=2cm] (isconf) {Conflict?};
    \node [block, below of=isconf, node distance=2cm] (conf) {Conflict Analysis};
    \node [block, left of=isconf, node distance=3.5cm, fill=lightgray, align=center] (progprop) {Programmatic \\ Propagation};
    \node [decision, left of=progprop, node distance=3.5cm, fill=lightgray, align=center, aspect=1.5] (isprogclause) {New Reason \\ Clauses?};
    \node [block, below of=isprogclause, node distance=2cm, fill=lightgray, align=center] (progconf) {Programmatic \\ Conflict Analysis};
    \node [decision, right of=progconf, node distance=3.5cm,fill=lightgray, align=center, aspect=1.5] (isconfclause) {New Conflict \\ Clauses?};
    \node [decision, right of=conf, node distance=3cm] (top) {Top Level?};
    \node [block, right of=unit, node distance=3cm] (backjump) {Backjump};
    \node [above right of=top, node distance=2cm] (unsat) {UNSAT};
    \node [decision, left of=isprogclause, node distance=3cm, align=center, aspect=1.5] (assigned) {All Variables \\ Assigned?};
    \node [block, above of=assigned, node distance=2cm] (decision) {Decision};
    \node [below left of=assigned, node distance=3cm] (sat) {SAT};
    
    \draw [->] (input) -- (unit);
    \draw [->] (unit) -- (isconf);
    \draw [->] (isconf) -- node {No} (progprop);
    \draw [->] (isconf) -- node {Yes} (conf);
    \draw [->] (progprop) -- (isprogclause);
    \draw [->] (isprogclause) -- node {No} (progconf);
    \draw [->] (isprogclause) |- node [pos=0.1] {Yes} (unit.190);
    \draw [->] (progconf) -- (isconfclause);
    \draw [->] (isconfclause) -- node {Yes} (conf);
    \draw [->] (isconfclause.south) -- ++(0,-5pt) -| node [pos=0.1] {No} (assigned);
    \draw [->] (assigned) -- node {No} (decision);
    \draw [->] (assigned) -| node [pos=0.2] {Yes} (sat);
    \draw [->] (decision.13) -- (unit.170);
    \draw [->] (backjump) -- (unit);
    \draw [->] (conf) -- (top);
    \draw [->] (top) -- node [pos=0.1] {No} (backjump);
    \draw [->] (top) -| node [pos=0.2] {Yes} (unsat);

    \end{tikzpicture}
    }

    \caption{Block Diagram of a CDCL SAT solver with the Programmatic components that implement
    cryptographic related reasoning (shaded blocks).}
    \label{fig:progsat}
    \Description{Block Diagram of a CDCL SAT solver with the Programmatic components (shaded blocks).}
\end{figure*}
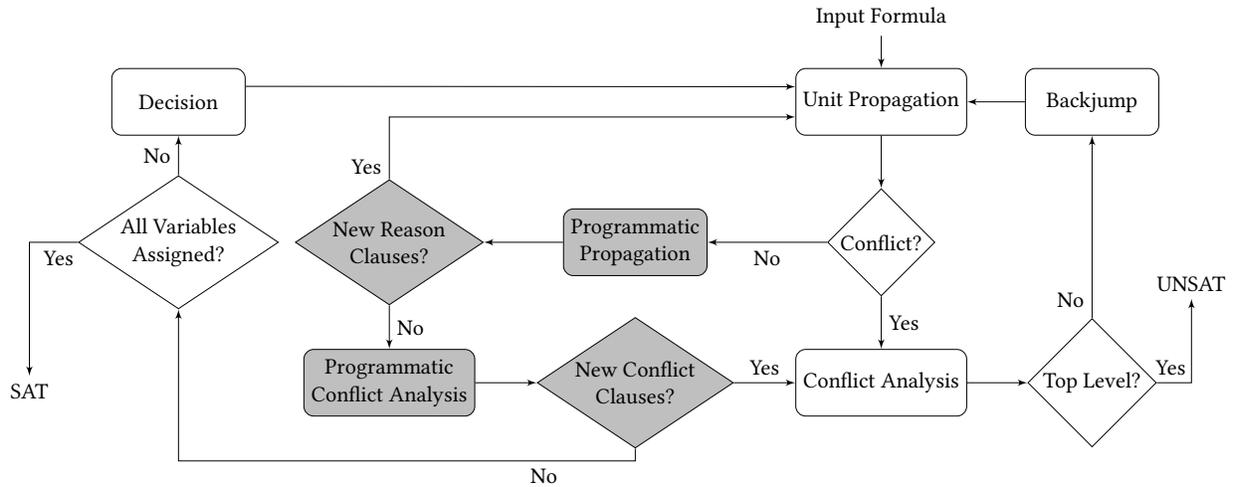

\section{\cdcl Framework} \label{sec:cdclcrypto}

In this section, we describe the \cdcl framework, based on a programmatic SAT solver, illustrated in Figure \ref{fig:progsat}.

\subsection{Programmatic Interface in SAT Solvers}
We call a SAT solver {\it programmatic}~\cite{ganesh2012lynx} if it is augmented with a set of callback functions that allow the user to add functionality to the solver's propagation and conflict analysis routines. The idea is inspired by the CDCL($T$) architecture, in which a theory solver provides support for theory propagation and theory conflict analysis to the base Boolean CDCL solver.  Programmatic SAT solving differs from the general concept in 3 ways: First, the theory solver in the context of programmatic SAT can be an arbitrary piece of code, in that we place no requirements on its completeness; second, this code might be particularized to every input to the solver. That is, unlike the $T$-solver in CDCL($T$) which remains invariant for all formulas from the language of $T$, the code added via the programmatic interface in a programmatic SAT solver can be specific and unique to each input; and finally, the interface of programmatic SAT solvers is much simpler than that of SMT solvers.

The main advantage of using programmatic SAT is that it allows easy customization of the SAT solver to specific Boolean instances rather than an entire theory. The developer thus has more fine-grained control over the power of the SAT solver. This architecture has also shown to be useful in solving problems in combinatorics \cite{bright2016mathcheck2}, and much more effective than only using a normal CNF encoding. Figure \ref{fig:progsat} shows the block diagram of a CDCL SAT solver and the connection of programmatic components (shaded blocks) to the main components.

\textbf{Programmatic propagation} has the role of providing clauses similar to theory propagation clauses. As can be seen in the figure, there is a close interaction loop between unit propagation and programmatic propagation, in which when the unit propagation is done, if there is no conflict, programmatic propagation analyzes the partial assignment and determines whether any other literal is implied according to the logic of the cryptographic function. If any literal is implied but missed by the unit propagation, an appropriate \textit{reason clause} is returned to empower the unit propagation. Consider that $\alpha$ is a subset of literals in the partial assignment that implies another literal $L$, and this implication is missed by unit propagation. The added reason clause will be simply $\alpha \rightarrow L$ (in CNF format). Then the unit propagation is invoked to set those literals and possibly find more implications that are caused by the new literals. Added reason clauses can be reused when the solver unassigns some of the variables and assigns them again (due to backjump or restart).

\textbf{Programmatic conflict analysis}, in a similar fashion, is invoked when the propagation is done (the combination of unit and programmatic) and no conflict is detected. It analyzes the partial assignment to check if there is conflicting information according to the domain knowledge. The user can return single or multiple conflict clauses if a conflict is detected. The core solver then looks at the variables that are in the conflict clause, and by examining the implication graph that has been built during the run of the solver, attempts to find a minimized root cause of the conflict.

We have implemented this framework on top of MapleSAT \cite{liang2016learning}. Programmatic routines need to know the mapping of the high level variables to the Boolean variable IDs. This is necessary to be able to verify the value of a predicate when the corresponding Boolean variables are set. To keep the variable ID mapping intact, we switched off the variable elimination procedure that MapleSAT performs as a preprocessing step to simplify the formula. During the search, the size of the conflict clause database only increases and this might negatively impact the performance of unit propagation. To handle this challenge, modern SAT solvers regularly delete some of the lower quality clauses. In the programmatic SAT, the same problem could happen for the reason clause database. In our implementation, we use the same clause deletion strategy of MapleSAT to prevent the overgrowth of reason clause database.

\subsection{Cryptographic Reasoning in Programmatic Callbacks}
Even for cryptographic functions that use very simple operations, like addition-rotation-xor (ARX) block ciphers and hash functions, some high level properties like commutativity of addition, is lost when translated into the Boolean level, let alone much more complex cryptographic properties. One can specifically encode these properties, but it will result in a very large SAT instance (e.g. commutativity of multi-operand additions in ARX). The programmatic approach enables us to express those properties concisely using a piece of code (C++ in our case), that are being used by the SAT solver through the programmatic interfaces. We will give more detailed use of these interfaces in two cryptanalysis applications. In section \ref{sec:afa}, we review an algebraic fault attack on SHA hash functions \cite{nejati2018algebraic} and present preliminary results on differential cryptanalysis of SHA-256 in section \ref{sec:diffcrypto}.


\section{Case Studies} \label{sec:cases}

\subsection{Algebraic Fault Attack} \label{sec:afa}

\begin{figure}[ht]
    \centering
    \includegraphics[width=0.45\textwidth]{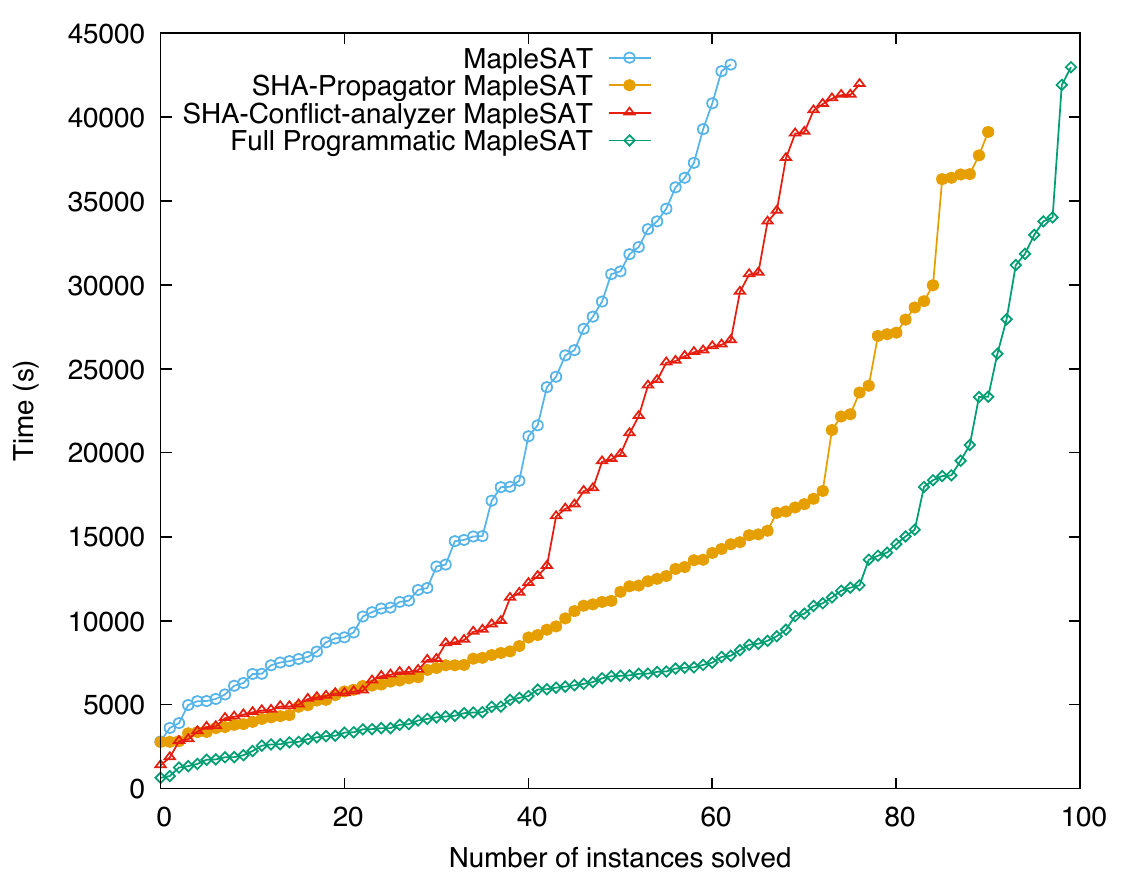}
    \caption{The cactus plot comparing MapleSAT with the MapleSAT after adding
	each of the programmatic callbacks on 32-bit fault attack on SHA-256 \cite{nejati2018algebraic}.
	Each data point ($X$, $Y$) on this plot
	means $X$ fault instances are solved under $Y$ seconds. Further down means solving faster
	and further right means solving more instances.}
    \label{fig:sha2cac}
\end{figure}

In this section, we review our enhancement of algebraic fault attack that has been applied to SHA-1 and SHA-256 using a programmatic SAT solver \cite{nejati2018algebraic} that enables us to solve AFA instances with much fewer number of injected faults. The solution verification loop is embedded in the programmatic conflict analysis. It has been observed in this work that when using the best performing encoding of the SHA function into SAT, if all of the input bits are set, although all the necessary information to derive output bits are available, unit propagation can not propagate the input values to output bits. On the other hand, when using an encoding that ensures the propagation of information to the output, the size of encoding becomes so large that the solver can not solve the instance in the given time limit. Therefore they proposed the use of programmatic propagation to keep the size of instance small and enhance the propagation. The programmatic components used in this work are briefly described as follows and their performance result is plotted in figure \ref{fig:sha2cac}.

\textbf{Programmatic Conflict Analysis}:
We are only interested in the values of variables that correspond to the secret message bits, which are a very small subset of all of the variables needed to encode the algebraic fault equation system into CNF. Whenever we solve the instance and find the message bits, we should check if it is a legitimate solution (hashes to the same correct hash output). Normally one could wait for the solver to finish solving the whole equation set and then check for the correctness, but we can do this verification as soon as the variables corresponding to the message bits are set. The sooner we reject a spurious solution, the faster the search process becomes. When the programmatic conflict analysis is invoked, first it recovers the original input message bits. If all message bit variables are set, it hashes the input message bits and checks it against the correct hash output. In case of mismatch, a conflict clause that blocks the current spurious message bits will be returned to the solver. The core does analysis using an implication graph on the returned conflict clauses and then goes through the backjumping procedure, as in the typical conflict analysis.

\textbf{Programmatic Propagation}:
It is mentioned in~\cite{philipp2015pblib} and~\cite{een2006translating} that encoding of a pseudo-Boolean constraint into CNF using adder networks, although providing a small and scalable instance, when running unit propagation over a partial assignment, might not find all of the implied literals that are implied in the original pseudo-Boolean constraint (or in the constraint satisfaction notation: does not maintain generalized arc-consistency). In SHA functions, we have multi-operand additions in each round. There are several encodings for these operations in the literature. Nossum's encoding \cite{nossum2012sat} gives a very compact CNF, which works very well in practice. But does not maintain the arc consistency. The programmatic propagation is called in the main search loop of the solver after unit propagation is done, and no conflicts are detected. The callback looks at the least significant bits of the operands in each of the multi-operand additions. If all bits up to some bit position $k$ are set, it checks if the $k$ least significant bits of the output are set as well. If they are not set, it returns clauses that encode the direct implication between input bits and output bit in the missing output bit positions. For an example of encoding implications, if $x=T$, $y=F$ is an assignment to the inputs of $z=x+y$ relation, and $z$ is not set, we return $x=T \land y=F \rightarrow z=T$ or $\neg x \lor y \lor z$. These implications force the solver to set the output bits in the next cycle. This ensures a directional (from input to output) consistency.

\subsection{Differential Cryptanalysis} \label{sec:diffcrypto}

A naive way of encoding an algebraic collision attack is to have two copies of a function $f$ that have constraints for having the same output and different inputs ($f(x) = f(y) \land x \neq y$). To improve upon this encoding, we can add a set of difference variables for all of the input, output and intermediate variables in the two copies, where each difference variable is the XOR of the two corresponding variables in the two copies. These difference variables are building the differential path. Just having the differential path does not make the problem easier, but by selecting a sparse differential path that is highly probable, the allowed combinations for variables in the two copies will reduce drastically. Note that for any operation, when we have \texttt{-} (no difference) in the input variables, we will have \texttt{-} at the output variables as well, i.e. running a function on the same input twice results in the same output. A sparse differential path means that most of the difference variables are forced to be \texttt{-}, and there should be few ``difference''s (\texttt{x}), to ensure different inputs and keep the possible combinations throughout the differential path limited. We put ``unknown'' (\texttt{?}) in the places that the effect of having difference in earlier steps can potentially be canceled (to be found by the solver). The common approach on differential cryptanalysis of hash functions is to find a differential path first (starting from a sparse path, find the values for \texttt{?}s), then use these constraints to find a conforming pair of messages that go through the two copies of the function that we had. There may be no pairs of messages that follow the path. In that case, we have to go back to the path and modify it. An important step in this process is the propagation of information throughout the differential path. In other words, having difference in the input of smaller operations, what is the possible set of combinations at the output of those operations (output difference). The implication from input differentials to output differentials is referred to as propagation rules.

Mendel et al. \cite{mendel2013improving} developed a dedicated tool for differential cryptanalysis of SHA-256. Prokop \cite{prokop2016differential} took their work and encoded their differential tables into SAT and studied the performance of different SAT solvers on them. Prokop shows collisions on SHA-256 up to 24 rounds, which is not matching the performance of Mendel's solver that gives a collision up to 31 rounds in the same attack model. Prokop is using bitwise XOR differences for encoding the difference possibilities. This means that he is using only \texttt{?} (unassigned), \texttt{-} and \texttt{x} values for a difference variable. The advantage of this approach is that each difference variable can be encoded with a single Boolean variable. But the disadvantage is that the propagation of information is less concrete in many cases. That is because a condition of for example \texttt{A} can not be expressed and thus it needs to fall back to the under-specified condition of \texttt{?}. To address this problem one can use multiple Boolean variables to encode each of the difference variables to cover all the possible information that is being propagated. The advantage of this approach is having more concrete possibilities and a more constrained set of values for pairs of message bits, but the disadvantage is that the instance becomes very large in terms of variables and clauses and the gain of having differential path constraints will be overshadowed by the complexity of the encoding. This is an opportunity for a programmatic component to implement the multi-valued logic of generalized conditions for difference variables while keeping the encoding of differential path simple. For example when using single Boolean variables, we can derive 2 propagation rules for the Boolean function $\text{IF}(x, y, z) = (x \land y) \lor (\neg x \land z)$, that are ``\texttt{-}\texttt{-}\texttt{-} $\rightarrow$ \texttt{-}'' and ``\texttt{-}\texttt{x}\texttt{x} $\rightarrow$ \texttt{x}'', and for the rest of input difference combinations, we can not imply any differential information for the output. But considering a multi-valued logic, we can have very fine-grained rules that rule out certain combinations for the pairs of bits at the output. Enumerating all of them gives us 1846 rules, which is expensive to encode in CNF.

In our implementation of programmatic propagation, simply put, we provide a truth table for each operation, that given input differences, determines and enforces the output difference if it is not \texttt{?}. Programmatic conflict analysis checks if the implied set of combinations of a difference variable does not have an intersection with a currently decided/deduced combination set. In other words, it looks whether after applying a propagation rule the difference variable becomes \texttt{\#}.

We took the differential path starting points from Prokop \cite{prokop2016differential}, but used our own encoding to translate the SHA-256 relations and differential path information into SAT\footnote{The encoder is available at: \url{https://github.com/saeednj/SAT-encoding}}. For encoding multi-operand addition we used Nossum's encoding \cite{nossum2012sat}. We ran MapleSAT (with and without the programmatic components) on these instances with a 24-hour time limit on Intel Core i7 CPU @ 3.4GHz and 16 GB of RAM. In table \ref{tab:difftiming}, MapleSAT(Crypto) refers to the version of MapleSAT that we instrumented with programmatic callbacks. As timings show, not only we can increase the number of rounds from 24 to 25, but also we can solve the instances of 25 rounds roughly 2.3 times faster when we use the programmatic interface.

\begin{table}[h]
    \caption{CPU times (in seconds) for SAT-based differential cryptanalysis (finding collisions) in 25 rounds of SHA-256.}
    \label{tab:difftiming}
    \begin{tabular}{|c|c|c|} \hline
        Solver & Encoding & Runtime (s) \\ \hline
        MapleSAT & Prokop\cite{prokop2016differential} & 29771.80 \\
        MapleSAT & Our encoding &  21926.60 \\
        MapleSAT(Crypto) & Our encoding & 12532.32 \\ \hline
    \end{tabular}
\end{table}

\section{Related Work}

Early works on the use of SAT solvers for cryptanalysis like finding cryptographic keys \cite{massacci1999using}, modular root finding \cite{fiorini2003fake}, or collision attack on MD5 \cite{mironov2006applications}, only used direct encoding of their problem to employ the power of SAT solvers. Subsequent works studied different ways of encoding the same problems into SAT to find formulas that are easier for a SAT solver in practice. Nossum \cite{nossum2012sat} and Morawiecki et al. \cite{morawiecki2013sat} presented instance generators for preimage attack on SHA-1 and SHA-3. 
To make the SAT-based attacks more powerful, De et al. \cite{de2007inversion} made use of Dobbertin's attack. They encoded the additional constraint alongside the main function to improve the base preimage attack on MD4. These types of cryptanalytic techniques can be encoded inside cryptographic reasoning components of the \cdcl to keep the size of instance small, but still have the benefit of reducing the size of search space.


The problem of finding the highest probable linear/differential trail has been studied for lightweight ciphers like Simon \cite{kolbl2015observations} and Speck \cite{ashur2017automated}. In these works, the task of finding the optimal trail is defined as an optimization problem, and at each step, an SMT solver (in particular STP \cite{ganesh2007decision}) is queried with a trail and a parameter. If the solver returns SAT the parameter is increased and the process is repeated until the optimal value is reached. 

Not all of the SAT-based cryptanalysis works have been completely blackbox. There were limited attempts to change the heuristics of the solver to improve the runtime. For example, Semenov et al. \cite{semenov2011parallel} changed the default activities and decay factor of VSIDS branching heuristics of Minisat and got better results. Although it should be mentioned that one can see this approach as configuring the parameters of the solver and not changing the algorithm. Prokop \cite{prokop2016differential} changed the branching heuristic of Lingeling to focus on the differential variables first in differential cryptanalysis of SHA-256. Furthermore, he studied value selection heuristics. For improving runtime of preimage attack on SHA-1 instances, an adaptive restart policy \cite{nejati2017adaptive} and a splitting heuristics for divide-and-conquer parallel SAT solvers \cite{nejati2017propagation} has been proposed.

Notable SAT-based tools that have been developed specifically for cryptanalysis (at least initially), include CryptLogVer \cite{morawiecki2013sat} and Transalg \cite{otpuschennikov2016encoding} which are tools for encoding cryptographic functions into SAT, CryptoMiniSat \cite{soos2009extending} which includes XOR reasoning, and CryptoSAT \cite{lafitte2014applications} and CryptoSMT \cite{CryptoSMT-ref} that provide higher level languages for expressing cryptographic relations. For solving the algebraic equation set of the cryptosystem, SAT and SMT solvers are usually used. But other types of solvers have also been shown beneficial. Mouha et al. \cite{mouha2011differential} use Mixed Integer Linear Programming solvers to find security boundaries in block ciphers.

Other than using off-the-shelf solvers, researchers have developed dedicated solvers to attack cryptographic primitives. Mostly these dedicated tools are based on guess-and-determine approach \cite{bard2009algebraic}, which is a method in algebraic cryptanalysis. In this method, we pick one variable with unknown value, guess a probable value for it, and then propagate the guessed information through the algebraic equation set that represents the cryptographic function, and in case of conflicting information, undo the guesses until the conflict is resolved. This is very similar to the process that a CDCL SAT solver follows (decision followed by unit propagation, and backtracking), but can be implemented specific to the function and not necessarily be in Boolean level. Mendel et al. \cite{mendel2011finding} developed a tool for differential cryptanalysis of SHA-256. They used random branching, problem specific propagation and backtracking. They improved their results by improving the search strategy, better local collisions and extra constraints \cite{mendel2013improving}. Eichlseder et al. \cite{eichlseder2014branching} took it further and improved the tool for SHA-512, by studying different branching heuristics. Although this tool is dedicated to this particular problem, it borrows many ideas from SAT solving. However, it is missing one of the most powerful components of a CDCL solver, which is conflict analysis. \cdcl has the potential to implement the higher level logic on the propagation of information, and at the same time, use the underlying conflict analysis of the core CDCL solver on the Boolean level representation of the relations.

\section{Conclusion}
We presented a framework for SAT-based cryptanalysis inspired by the CDCL($T$) paradigm. \cdcl consists of a core Boolean SAT solver that is instrumented with programmatic callbacks for propagation and conflict analysis. These callbacks will contain user-provided cryptographic reasoning, similar to a $T$-solver in CDCL($T$). This framework helps to have the higher level semantics of the cryptographic primitive available while keeping the size of the encoded function into SAT small and practical for the core SAT solver. \cdcl enables the researchers to implement their cryptanalytic techniques on top of a powerful search engine. This framework has been applied to algebraic fault analysis of SHA cryptographic hash functions and resulted in a much more effective search that requires far fewer number of injected faults. Also, a work in progress on the application of this framework on differential cryptanalysis has been demonstrated in this paper, which improves the number of rounds and the runtime of finding a collision for a round-reduced version of SHA-256 with 25 rounds. Symmetric cryptographic function designers usually test their designs against known attacks and cryptanalysis techniques. Automating these techniques helps with speeding up the design cycle. We believe that this framework has a great potential for improving the blackbox SAT-based cryptanalysis and therefore a valuable step toward automating cryptanalysis of cryptographic primitives.



\bibliographystyle{ACM-Reference-Format}
\bibliography{main}


\begin{thebibliography}{37}


\ifx \showCODEN    \undefined \def \showCODEN     #1{\unskip}     \fi
\ifx \showDOI      \undefined \def \showDOI       #1{#1}\fi
\ifx \showISBNx    \undefined \def \showISBNx     #1{\unskip}     \fi
\ifx \showISBNxiii \undefined \def \showISBNxiii  #1{\unskip}     \fi
\ifx \showISSN     \undefined \def \showISSN      #1{\unskip}     \fi
\ifx \showLCCN     \undefined \def \showLCCN      #1{\unskip}     \fi
\ifx \shownote     \undefined \def \shownote      #1{#1}          \fi
\ifx \showarticletitle \undefined \def \showarticletitle #1{#1}   \fi
\ifx \showURL      \undefined \def \showURL       {\relax}        \fi
\providecommand\bibfield[2]{#2}
\providecommand\bibinfo[2]{#2}
\providecommand\natexlab[1]{#1}
\providecommand\showeprint[2][]{arXiv:#2}

\bibitem[\protect\citeauthoryear{Ashur, De~Witte, and Liu}{Ashur
  et~al\mbox{.}}{2017}]%
        {ashur2017automated}
\bibfield{author}{\bibinfo{person}{Tomer Ashur}, \bibinfo{person}{Glenn
  De~Witte}, {and} \bibinfo{person}{Yunwen Liu}.}
  \bibinfo{year}{2017}\natexlab{}.
\newblock \showarticletitle{An Automated Tool for Rotational-XOR Cryptanalysis
  of ARX-based Primitives}. In \bibinfo{booktitle}{\emph{Proceedings of the
  38th Symposium on Information Theory in the Benelux}}. Werkgemeenschap voor
  Informatie-en Communicatietheorie, \bibinfo{pages}{59--66}.
\newblock


\bibitem[\protect\citeauthoryear{Audemard and Simon}{Audemard and
  Simon}{2009}]%
        {audemard2009predicting}
\bibfield{author}{\bibinfo{person}{Gilles Audemard} {and}
  \bibinfo{person}{Laurent Simon}.} \bibinfo{year}{2009}\natexlab{}.
\newblock \showarticletitle{{P}redicting learnt clauses quality in modern {SAT}
  solvers}. In \bibinfo{booktitle}{\emph{IJCAI}}, Vol.~\bibinfo{volume}{9}.
  \bibinfo{pages}{399--404}.
\newblock


\bibitem[\protect\citeauthoryear{Audemard and Simon}{Audemard and
  Simon}{2012}]%
        {audemard2012refining}
\bibfield{author}{\bibinfo{person}{Gilles Audemard} {and}
  \bibinfo{person}{Laurent Simon}.} \bibinfo{year}{2012}\natexlab{}.
\newblock \showarticletitle{{R}efining {R}estarts {S}trategies for {SAT} and
  {UNSAT}}. In \bibinfo{booktitle}{\emph{Principles and Practice of Constraint
  Programming}}. Springer, \bibinfo{pages}{118--126}.
\newblock


\bibitem[\protect\citeauthoryear{Bard}{Bard}{2009}]%
        {bard2009algebraic}
\bibfield{author}{\bibinfo{person}{Gregory Bard}.}
  \bibinfo{year}{2009}\natexlab{}.
\newblock \bibinfo{booktitle}{\emph{Algebraic cryptanalysis}}.
\newblock \bibinfo{publisher}{Springer Science \& Business Media}.
\newblock


\bibitem[\protect\citeauthoryear{Biere, Heule, and van Maaren}{Biere
  et~al\mbox{.}}{2009}]%
        {biere2009handbook}
\bibfield{author}{\bibinfo{person}{Armin Biere}, \bibinfo{person}{Marijn
  Heule}, {and} \bibinfo{person}{Hans van Maaren}.}
  \bibinfo{year}{2009}\natexlab{}.
\newblock \bibinfo{booktitle}{\emph{Handbook of satisfiability}}.
  Vol.~\bibinfo{volume}{185}.
\newblock \bibinfo{publisher}{IOS press}.
\newblock


\bibitem[\protect\citeauthoryear{Biham and Shamir}{Biham and Shamir}{1991}]%
        {biham1991differential}
\bibfield{author}{\bibinfo{person}{Eli Biham} {and} \bibinfo{person}{Adi
  Shamir}.} \bibinfo{year}{1991}\natexlab{}.
\newblock \showarticletitle{Differential cryptanalysis of DES-like
  cryptosystems}.
\newblock \bibinfo{journal}{\emph{Journal of CRYPTOLOGY}} \bibinfo{volume}{4},
  \bibinfo{number}{1} (\bibinfo{year}{1991}), \bibinfo{pages}{3--72}.
\newblock


\bibitem[\protect\citeauthoryear{Bright, Ganesh, Heinle, Kotsireas, Nejati, and
  Czarnecki}{Bright et~al\mbox{.}}{2016}]%
        {bright2016mathcheck2}
\bibfield{author}{\bibinfo{person}{Curtis Bright}, \bibinfo{person}{Vijay
  Ganesh}, \bibinfo{person}{Albert Heinle}, \bibinfo{person}{Ilias Kotsireas},
  \bibinfo{person}{Saeed Nejati}, {and} \bibinfo{person}{Krzysztof Czarnecki}.}
  \bibinfo{year}{2016}\natexlab{}.
\newblock \showarticletitle{MathCheck2: A {SAT}+ {CAS} Verifier for
  Combinatorial Conjectures}. In \bibinfo{booktitle}{\emph{International
  Workshop on Computer Algebra in Scientific Computing}}. Springer,
  \bibinfo{pages}{117--133}.
\newblock


\bibitem[\protect\citeauthoryear{Cadar, Ganesh, Pawlowski, Dill, and
  Engler}{Cadar et~al\mbox{.}}{2008}]%
        {cadar2008exe}
\bibfield{author}{\bibinfo{person}{Cristian Cadar}, \bibinfo{person}{Vijay
  Ganesh}, \bibinfo{person}{Peter~M Pawlowski}, \bibinfo{person}{David~L Dill},
  {and} \bibinfo{person}{Dawson~R Engler}.} \bibinfo{year}{2008}\natexlab{}.
\newblock \showarticletitle{{EXE}: automatically generating inputs of death}.
\newblock \bibinfo{journal}{\emph{ACM Transactions on Information and System
  Security (TISSEC)}} \bibinfo{volume}{12}, \bibinfo{number}{2}
  (\bibinfo{year}{2008}), \bibinfo{pages}{10}.
\newblock


\bibitem[\protect\citeauthoryear{De, Kumarasubramanian, and Venkatesan}{De
  et~al\mbox{.}}{2007}]%
        {de2007inversion}
\bibfield{author}{\bibinfo{person}{Debapratim De}, \bibinfo{person}{Abishek
  Kumarasubramanian}, {and} \bibinfo{person}{Ramarathnam Venkatesan}.}
  \bibinfo{year}{2007}\natexlab{}.
\newblock \showarticletitle{Inversion attacks on secure hash functions using
  {SAT} solvers}. In \bibinfo{booktitle}{\emph{International Conference on
  Theory and Applications of Satisfiability Testing}}. Springer,
  \bibinfo{pages}{377--382}.
\newblock


\bibitem[\protect\citeauthoryear{De~Canniere and Rechberger}{De~Canniere and
  Rechberger}{2006}]%
        {de2006finding}
\bibfield{author}{\bibinfo{person}{Christophe De~Canniere} {and}
  \bibinfo{person}{Christian Rechberger}.} \bibinfo{year}{2006}\natexlab{}.
\newblock \showarticletitle{{F}inding {SHA}-1 characteristics: general results
  and applications}.
\newblock In \bibinfo{booktitle}{\emph{Advances in Cryptology--ASIACRYPT
  2006}}. \bibinfo{publisher}{Springer}, \bibinfo{pages}{1--20}.
\newblock


\bibitem[\protect\citeauthoryear{E{\'e}n and Sorensson}{E{\'e}n and
  Sorensson}{2006}]%
        {een2006translating}
\bibfield{author}{\bibinfo{person}{Niklas E{\'e}n} {and}
  \bibinfo{person}{Niklas Sorensson}.} \bibinfo{year}{2006}\natexlab{}.
\newblock \showarticletitle{Translating pseudo-boolean constraints into {SAT}}.
\newblock \bibinfo{journal}{\emph{Journal on Satisfiability, Boolean Modeling
  and Computation}}  \bibinfo{volume}{2} (\bibinfo{year}{2006}),
  \bibinfo{pages}{1--26}.
\newblock


\bibitem[\protect\citeauthoryear{Eichlseder, Mendel, and
  Schl{\"a}ffer}{Eichlseder et~al\mbox{.}}{2014}]%
        {eichlseder2014branching}
\bibfield{author}{\bibinfo{person}{Maria Eichlseder}, \bibinfo{person}{Florian
  Mendel}, {and} \bibinfo{person}{Martin Schl{\"a}ffer}.}
  \bibinfo{year}{2014}\natexlab{}.
\newblock \showarticletitle{Branching heuristics in differential collision
  search with applications to SHA-512}. In
  \bibinfo{booktitle}{\emph{International Workshop on Fast Software
  Encryption}}. Springer, \bibinfo{pages}{473--488}.
\newblock


\bibitem[\protect\citeauthoryear{Fiorini, Martinelli, and Massacci}{Fiorini
  et~al\mbox{.}}{2003}]%
        {fiorini2003fake}
\bibfield{author}{\bibinfo{person}{Claudia Fiorini}, \bibinfo{person}{Enrico
  Martinelli}, {and} \bibinfo{person}{Fabio Massacci}.}
  \bibinfo{year}{2003}\natexlab{}.
\newblock \showarticletitle{How to fake an RSA signature by encoding modular
  root finding as a SAT problem}.
\newblock \bibinfo{journal}{\emph{Discrete Applied Mathematics}}
  \bibinfo{volume}{130}, \bibinfo{number}{2} (\bibinfo{year}{2003}),
  \bibinfo{pages}{101--127}.
\newblock


\bibitem[\protect\citeauthoryear{Ganesh and Dill}{Ganesh and Dill}{2007}]%
        {ganesh2007decision}
\bibfield{author}{\bibinfo{person}{Vijay Ganesh} {and} \bibinfo{person}{David~L
  Dill}.} \bibinfo{year}{2007}\natexlab{}.
\newblock \showarticletitle{A decision procedure for bit-vectors and arrays}.
  In \bibinfo{booktitle}{\emph{International Conference on Computer Aided
  Verification}}. Springer, \bibinfo{pages}{519--531}.
\newblock


\bibitem[\protect\citeauthoryear{Ganesh, O'Donnell, Soos, Devadas, Rinard, and
  Solar{-}Lezama}{Ganesh et~al\mbox{.}}{2012}]%
        {ganesh2012lynx}
\bibfield{author}{\bibinfo{person}{Vijay Ganesh}, \bibinfo{person}{Charles~W.
  O'Donnell}, \bibinfo{person}{Mate Soos}, \bibinfo{person}{Srinivas Devadas},
  \bibinfo{person}{Martin~C. Rinard}, {and} \bibinfo{person}{Armando
  Solar{-}Lezama}.} \bibinfo{year}{2012}\natexlab{}.
\newblock \showarticletitle{Lynx: {A} Programmatic {SAT} Solver for the
  {RNA}-Folding Problem}. In \bibinfo{booktitle}{\emph{Theory and Applications
  of Satisfiability Testing - {SAT} 2012 - 15th International Conference,
  Trento, Italy, June 17-20, 2012. Proceedings}}. \bibinfo{pages}{143--156}.
\newblock


\bibitem[\protect\citeauthoryear{K{\"o}lbl, Leander, and Tiessen}{K{\"o}lbl
  et~al\mbox{.}}{2015}]%
        {kolbl2015observations}
\bibfield{author}{\bibinfo{person}{Stefan K{\"o}lbl}, \bibinfo{person}{Gregor
  Leander}, {and} \bibinfo{person}{Tyge Tiessen}.}
  \bibinfo{year}{2015}\natexlab{}.
\newblock \showarticletitle{Observations on the SIMON block cipher family}. In
  \bibinfo{booktitle}{\emph{Annual Cryptology Conference}}. Springer,
  \bibinfo{pages}{161--185}.
\newblock


\bibitem[\protect\citeauthoryear{Lafitte, Nakahara~Jr, and Van~Heule}{Lafitte
  et~al\mbox{.}}{2014}]%
        {lafitte2014applications}
\bibfield{author}{\bibinfo{person}{Fr{\'e}d{\'e}ric Lafitte},
  \bibinfo{person}{Jorge Nakahara~Jr}, {and} \bibinfo{person}{Dirk Van~Heule}.}
  \bibinfo{year}{2014}\natexlab{}.
\newblock \showarticletitle{Applications of SAT solvers in cryptanalysis:
  finding weak keys and preimages}.
\newblock \bibinfo{journal}{\emph{Journal on Satisfiability, Boolean Modeling
  and Computation}}  \bibinfo{volume}{9} (\bibinfo{year}{2014}),
  \bibinfo{pages}{1--25}.
\newblock


\bibitem[\protect\citeauthoryear{Liang, Ganesh, Poupart, and Czarnecki}{Liang
  et~al\mbox{.}}{2016}]%
        {liang2016learning}
\bibfield{author}{\bibinfo{person}{Jia~Hui Liang}, \bibinfo{person}{Vijay
  Ganesh}, \bibinfo{person}{Pascal Poupart}, {and} \bibinfo{person}{Krzysztof
  Czarnecki}.} \bibinfo{year}{2016}\natexlab{}.
\newblock \showarticletitle{Learning rate based branching heuristic for SAT
  solvers}. In \bibinfo{booktitle}{\emph{International Conference on Theory and
  Applications of Satisfiability Testing}}. Springer International Publishing,
  \bibinfo{pages}{123--140}.
\newblock


\bibitem[\protect\citeauthoryear{Marques-Silva and Sakallah}{Marques-Silva and
  Sakallah}{1999}]%
        {marques1999grasp}
\bibfield{author}{\bibinfo{person}{Jo{\~a}o~P Marques-Silva} {and}
  \bibinfo{person}{Karem~A Sakallah}.} \bibinfo{year}{1999}\natexlab{}.
\newblock \showarticletitle{{GRASP}: a search algorithm for propositional
  satisfiability}.
\newblock \bibinfo{journal}{\emph{Computers, IEEE Transactions on}}
  \bibinfo{volume}{48}, \bibinfo{number}{5} (\bibinfo{year}{1999}),
  \bibinfo{pages}{506--521}.
\newblock


\bibitem[\protect\citeauthoryear{Massacci}{Massacci}{1999}]%
        {massacci1999using}
\bibfield{author}{\bibinfo{person}{Fabio Massacci}.}
  \bibinfo{year}{1999}\natexlab{}.
\newblock \showarticletitle{Using Walk-SAT and Rel-SAT for cryptographic key
  search}. In \bibinfo{booktitle}{\emph{IJCAI}}, Vol.~\bibinfo{volume}{1999}.
  \bibinfo{pages}{290--295}.
\newblock


\bibitem[\protect\citeauthoryear{Mendel, Nad, and Schl{\"a}ffer}{Mendel
  et~al\mbox{.}}{2011}]%
        {mendel2011finding}
\bibfield{author}{\bibinfo{person}{Florian Mendel}, \bibinfo{person}{Tomislav
  Nad}, {and} \bibinfo{person}{Martin Schl{\"a}ffer}.}
  \bibinfo{year}{2011}\natexlab{}.
\newblock \showarticletitle{Finding SHA-2 characteristics: searching through a
  minefield of contradictions}. In \bibinfo{booktitle}{\emph{International
  Conference on the Theory and Application of Cryptology and Information
  Security}}. Springer, \bibinfo{pages}{288--307}.
\newblock


\bibitem[\protect\citeauthoryear{Mendel, Nad, and Schl{\"a}ffer}{Mendel
  et~al\mbox{.}}{2013}]%
        {mendel2013improving}
\bibfield{author}{\bibinfo{person}{Florian Mendel}, \bibinfo{person}{Tomislav
  Nad}, {and} \bibinfo{person}{Martin Schl{\"a}ffer}.}
  \bibinfo{year}{2013}\natexlab{}.
\newblock \showarticletitle{Improving local collisions: new attacks on reduced
  SHA-256}. In \bibinfo{booktitle}{\emph{Annual International Conference on the
  Theory and Applications of Cryptographic Techniques}}. Springer,
  \bibinfo{pages}{262--278}.
\newblock


\bibitem[\protect\citeauthoryear{Mironov and Zhang}{Mironov and Zhang}{2006}]%
        {mironov2006applications}
\bibfield{author}{\bibinfo{person}{Ilya Mironov} {and} \bibinfo{person}{Lintao
  Zhang}.} \bibinfo{year}{2006}\natexlab{}.
\newblock \showarticletitle{Applications of SAT solvers to cryptanalysis of
  hash functions}. In \bibinfo{booktitle}{\emph{International Conference on
  Theory and Applications of Satisfiability Testing}}. Springer,
  \bibinfo{pages}{102--115}.
\newblock


\bibitem[\protect\citeauthoryear{Morawiecki and Srebrny}{Morawiecki and
  Srebrny}{2013}]%
        {morawiecki2013sat}
\bibfield{author}{\bibinfo{person}{Pawe{\l} Morawiecki} {and}
  \bibinfo{person}{Marian Srebrny}.} \bibinfo{year}{2013}\natexlab{}.
\newblock \showarticletitle{A SAT-based preimage analysis of reduced KECCAK
  hash functions}.
\newblock \bibinfo{journal}{\emph{Inform. Process. Lett.}}
  \bibinfo{volume}{113}, \bibinfo{number}{10-11} (\bibinfo{year}{2013}),
  \bibinfo{pages}{392--397}.
\newblock


\bibitem[\protect\citeauthoryear{Moskewicz, Madigan, Zhao, Zhang, and
  Malik}{Moskewicz et~al\mbox{.}}{2001}]%
        {moskewicz2001chaff}
\bibfield{author}{\bibinfo{person}{Matthew~W Moskewicz},
  \bibinfo{person}{Conor~F Madigan}, \bibinfo{person}{Ying Zhao},
  \bibinfo{person}{Lintao Zhang}, {and} \bibinfo{person}{Sharad Malik}.}
  \bibinfo{year}{2001}\natexlab{}.
\newblock \showarticletitle{{C}haff: engineering an efficient {SAT} solver}. In
  \bibinfo{booktitle}{\emph{Proceedings of the 38th annual Design Automation
  Conference}}. ACM, \bibinfo{pages}{530--535}.
\newblock


\bibitem[\protect\citeauthoryear{Mouha, Wang, Gu, and Preneel}{Mouha
  et~al\mbox{.}}{2011}]%
        {mouha2011differential}
\bibfield{author}{\bibinfo{person}{Nicky Mouha}, \bibinfo{person}{Qingju Wang},
  \bibinfo{person}{Dawu Gu}, {and} \bibinfo{person}{Bart Preneel}.}
  \bibinfo{year}{2011}\natexlab{}.
\newblock \showarticletitle{Differential and linear cryptanalysis using
  mixed-integer linear programming}. In \bibinfo{booktitle}{\emph{International
  Conference on Information Security and Cryptology}}. Springer,
  \bibinfo{pages}{57--76}.
\newblock


\bibitem[\protect\citeauthoryear{Nejati, Hor{\'a}{\v{c}}ek, Gebotys, and
  Ganesh}{Nejati et~al\mbox{.}}{2018}]%
        {nejati2018algebraic}
\bibfield{author}{\bibinfo{person}{Saeed Nejati}, \bibinfo{person}{Jan
  Hor{\'a}{\v{c}}ek}, \bibinfo{person}{Catherine Gebotys}, {and}
  \bibinfo{person}{Vijay Ganesh}.} \bibinfo{year}{2018}\natexlab{}.
\newblock \showarticletitle{Algebraic Fault Attack on SHA Hash Functions Using
  Programmatic SAT Solvers}. In \bibinfo{booktitle}{\emph{International
  Conference on Principles and Practice of Constraint Programming}}. Springer,
  \bibinfo{pages}{737--754}.
\newblock


\bibitem[\protect\citeauthoryear{Nejati, Liang, Gebotys, Czarnecki, and
  Ganesh}{Nejati et~al\mbox{.}}{2017a}]%
        {nejati2017adaptive}
\bibfield{author}{\bibinfo{person}{Saeed Nejati}, \bibinfo{person}{Jia~Hui
  Liang}, \bibinfo{person}{Catherine Gebotys}, \bibinfo{person}{Krzysztof
  Czarnecki}, {and} \bibinfo{person}{Vijay Ganesh}.}
  \bibinfo{year}{2017}\natexlab{a}.
\newblock \showarticletitle{Adaptive Restart and {CEGAR}-based Solver for
  Inverting Cryptographic Hash Functions}. In \bibinfo{booktitle}{\emph{Working
  Conference on Verified Software: Theories, Tools, and Experiments}}.
  Springer, \bibinfo{pages}{120--131}.
\newblock


\bibitem[\protect\citeauthoryear{Nejati, Newsham, Scott, Liang, Gebotys,
  Poupart, and Ganesh}{Nejati et~al\mbox{.}}{2017b}]%
        {nejati2017propagation}
\bibfield{author}{\bibinfo{person}{Saeed Nejati}, \bibinfo{person}{Zack
  Newsham}, \bibinfo{person}{Joseph Scott}, \bibinfo{person}{Jia~Hui Liang},
  \bibinfo{person}{Catherine Gebotys}, \bibinfo{person}{Pascal Poupart}, {and}
  \bibinfo{person}{Vijay Ganesh}.} \bibinfo{year}{2017}\natexlab{b}.
\newblock \showarticletitle{A propagation rate based splitting heuristic for
  divide-and-conquer solvers}. In \bibinfo{booktitle}{\emph{International
  Conference on Theory and Applications of Satisfiability Testing}}. Springer,
  \bibinfo{pages}{251--260}.
\newblock


\bibitem[\protect\citeauthoryear{Nossum}{Nossum}{2012}]%
        {nossum2012sat}
\bibfield{author}{\bibinfo{person}{Vegard Nossum}.}
  \bibinfo{year}{2012}\natexlab{}.
\newblock \showarticletitle{{SAT}-based {P}reimage {A}ttacks on {SHA}-1}.
\newblock  (\bibinfo{year}{2012}).
\newblock


\bibitem[\protect\citeauthoryear{Otpuschennikov, Semenov, Gribanova, Zaikin,
  and Kochemazov}{Otpuschennikov et~al\mbox{.}}{2016}]%
        {otpuschennikov2016encoding}
\bibfield{author}{\bibinfo{person}{Ilya Otpuschennikov},
  \bibinfo{person}{Alexander Semenov}, \bibinfo{person}{Irina Gribanova},
  \bibinfo{person}{Oleg Zaikin}, {and} \bibinfo{person}{Stepan Kochemazov}.}
  \bibinfo{year}{2016}\natexlab{}.
\newblock \showarticletitle{Encoding cryptographic functions to SAT using
  Transalg system}. In \bibinfo{booktitle}{\emph{Proceedings of the
  Twenty-second European Conference on Artificial Intelligence}}. IOS Press,
  \bibinfo{pages}{1594--1595}.
\newblock


\bibitem[\protect\citeauthoryear{Philipp and Steinke}{Philipp and
  Steinke}{2015}]%
        {philipp2015pblib}
\bibfield{author}{\bibinfo{person}{Tobias Philipp} {and} \bibinfo{person}{Peter
  Steinke}.} \bibinfo{year}{2015}\natexlab{}.
\newblock \showarticletitle{PBLib: a library for encoding pseudo-boolean
  constraints into {CNF}}. In \bibinfo{booktitle}{\emph{International
  Conference on Theory and Applications of Satisfiability Testing}}. Springer,
  \bibinfo{pages}{9--16}.
\newblock


\bibitem[\protect\citeauthoryear{Prokop}{Prokop}{2016}]%
        {prokop2016differential}
\bibfield{author}{\bibinfo{person}{Lukas Prokop}.}
  \bibinfo{year}{2016}\natexlab{}.
\newblock \emph{\bibinfo{title}{Differential cryptanalysis with SAT solvers}}.
\newblock \bibinfo{thesistype}{Ph.D. Dissertation}. \bibinfo{school}{University
  of Technology, Graz}.
\newblock


\bibitem[\protect\citeauthoryear{Rintanen}{Rintanen}{2009}]%
        {rintanen2009planning}
\bibfield{author}{\bibinfo{person}{Jussi Rintanen}.}
  \bibinfo{year}{2009}\natexlab{}.
\newblock \showarticletitle{{P}lanning and {SAT}.}
\newblock \bibinfo{journal}{\emph{Handbook of Satisfiability}}
  \bibinfo{volume}{185} (\bibinfo{year}{2009}), \bibinfo{pages}{483--504}.
\newblock


\bibitem[\protect\citeauthoryear{Semenov, Zaikin, Bespalov, and
  Posypkin}{Semenov et~al\mbox{.}}{2011}]%
        {semenov2011parallel}
\bibfield{author}{\bibinfo{person}{Alexander Semenov}, \bibinfo{person}{Oleg
  Zaikin}, \bibinfo{person}{Dmitry Bespalov}, {and} \bibinfo{person}{Mikhail
  Posypkin}.} \bibinfo{year}{2011}\natexlab{}.
\newblock \showarticletitle{Parallel logical cryptanalysis of the generator
  A5/1 in BNB-Grid system}. In \bibinfo{booktitle}{\emph{International
  Conference on Parallel Computing Technologies}}. Springer,
  \bibinfo{pages}{473--483}.
\newblock


\bibitem[\protect\citeauthoryear{Soos, Nohl, and Castelluccia}{Soos
  et~al\mbox{.}}{2009}]%
        {soos2009extending}
\bibfield{author}{\bibinfo{person}{Mate Soos}, \bibinfo{person}{Karsten Nohl},
  {and} \bibinfo{person}{Claude Castelluccia}.}
  \bibinfo{year}{2009}\natexlab{}.
\newblock \showarticletitle{Extending SAT solvers to cryptographic problems}.
  In \bibinfo{booktitle}{\emph{International Conference on Theory and
  Applications of Satisfiability Testing}}. Springer,
  \bibinfo{pages}{244--257}.
\newblock


\bibitem[\protect\citeauthoryear{{Stefan K{\"o}lbl}}{{Stefan
  K{\"o}lbl}}{[n.d.]}]%
        {CryptoSMT-ref}
\bibfield{author}{\bibinfo{person}{{Stefan K{\"o}lbl}}.}
  \bibinfo{year}{[n.d.]}\natexlab{}.
\newblock \bibinfo{title}{{CryptoSMT: An easy to use tool for cryptanalysis of
  symmetric primitives}}.
\newblock
\newblock
\newblock
\shownote{\url{https://github.com/kste/cryptosmt}.}


\end{thebibliography}

\end{document}